\documentclass[pra,aps,twocolumn,10pt,showpacs,groupedaddress,superscriptaddress,floatfix,notitlepage]{revtex4-2}

\usepackage{amsmath,amsfonts,amssymb,amsthm,graphics,graphicx,epsfig,color,times, braket}
\usepackage{graphicx}
\usepackage{dcolumn}
\usepackage{bm}
\usepackage[utf8x]{inputenc}
\usepackage{color}
\usepackage{bbm, dsfont}
\usepackage{subfigure}
\usepackage{hyperref}
\hypersetup{
	colorlinks   = true, 
	urlcolor     = blue,
	linkcolor    = blue, 
	citecolor   = blue 
}
\usepackage{mathrsfs}
\usepackage{verbatim}
\usepackage{natbib}
\usepackage{physics}
\usepackage{booktabs}
\usepackage{float}
\usepackage{enumerate}
\ifdefined\directlua
\usepackage{fontspec}
\else
\usepackage[T1]{fontenc}
\usepackage[nomath]{lmodern}
\fi

\begin{document}


\title{\textbf{Strongly Confined Atomic Excitation Localization in a Weakly-Driven Atom-Waveguide Interface}}

\author{Shao-Hung Chung}
\email{ryan2031823@gmail.com}
\affiliation{Institute of Atomic and Molecular Sciences, Academia Sinica, Taipei 10617, Taiwan}

\author{Wei Chen}
\affiliation{Institute of Atomic and Molecular Sciences, Academia Sinica, Taipei 10617, Taiwan}
\affiliation{Department of Physics and Center for Quantum Science and Engineering, National Taiwan University, Taipei 10617, Taiwan}

\author{H. H. Jen}%
\affiliation{Institute of Atomic and Molecular Sciences, Academia Sinica, Taipei 10617, Taiwan}
\affiliation{Molecular Science and Technology Program, Taiwan International Graduate Program, Academia Sinica, Taiwan}
\affiliation{Physics Division, National Center for Theoretical Sciences, Taipei 10617, Taiwan}

\date{\today}

\begin{abstract}
An atomic array coupled to a photonic crystal waveguide forms a strongly coupled quantum interface, exhibiting various intriguing collective features of quantum dynamics. Here we consider a homogeneous atomic array and theoretically investigate its steady-state distribution when the incident fields drive the atoms from both sides at asymmetric angles. This effectively creates an interface shared by two zones of atoms under different driving angles. This setup introduces a competition between photon-mediated dipole-dipole interactions and the directionality of coupling, while differences of the travelling phases from the incident angles further influence the overall steady-state behavior. Under this asymmetric driving scheme, the presence of strongly confined localization can be identified, where localization can occur either at the interface or at one of edges. Additionally, we examine the size effect on the atomic localization, deriving an empirical formula to predict parameter regimes that favor interfaced localization. We also consider a defect-driving scheme, where a third zone is created by  undriven atoms under symmetric travelling phases. This results in strongly confined single-site excitation localization, which can be explained through analytical solutions under the reciprocal coupling. Finally, we propose several methods for precise control of multiple single-site localizations under the defect-driving scheme. Our results provide insights into driven-dissipative quantum systems with nonreciprocal couplings and pave the way for quantum simulation of exotic many-body states relevant to quantum information applications.
\end{abstract}

\maketitle

\section{Introduction}
\label{sec:introduction}

Chiral-coupled atomic systems constitute a compelling platform in atom-waveguide quantum electrodynamics \cite{RevModPhys.95.015002, Chiralquantumoptics, RevModPhys.90.031002, PhysRevResearch.2.043213, PhysRevX.11.011015, PhysRevResearch.3.033233,suárezforero2024chiralquantumopticsrecent,jen2024photonmediateddipoledipoleinteractionsresource}, enabling engineering over the directionality of light propagation. This breaks the time-reversal symmetry in light-matter couplings and establishes effective nonreciprocal decay channels. For instance, unidirectional coupling can be achieved through spin-momentum locking \cite{BLIOKH20151}, while these decay channels can be tuned by applying external magnetic fields \cite{Chiralquantumoptics,suárezforero2024chiralquantumopticsrecent,mitsch2014quantum,PhysRevLett.113.237203,PhysRevA.91.042116}, thereby allowing control of light transmission based on the quantum emitter's internal state. Such controlled and highly efficient nonreciprocal coupling between atoms has been demonstrated in strongly coupled systems, including artificial quantum emitters \cite{PhysRevLett.110.037402,PhysRevLett.113.093603,PhysRevLett.113.143601,Deterministic,synthetic-magnetic-field,Synthesis-of-antisymmetric-spin-exchange}, atom-nanofiber \cite{Corzo2019,Solano2017,mitsch2014quantum,Sayrin-15}, atom-photonic crystal waveguide \cite{PCW1,PhysRevA.104.053703}, and diamond nanophotonic platforms \cite{doi:10.1126/science.aah6875,PhysRevLett.118.223603}.

Under external laser driving, these platforms form distinctive driven-dissipative open quantum systems \cite{PhysRevLett.104.203603,Corzo2019,Kim2019,doi:10.1126/science.1237125,PhysRevLett.115.063601,PhysRevA.102.043525,doi:10.1126/science.abi9917,PhysRevResearch.4.023002,PhysRevA.106.043723}. The interplay between dissipation and interaction strengths facilitates the emergence of novel quantum many-body states and a wealth of dynamical phenomena. Examples include nontrivial collective radiation \cite{PhysRevA.99.023802,PhysRevLett.122.203605,PhysRevLett.123.253601,Albrecht_2019,PhysRevA.101.023830,PhysRevX.10.031011,PhysRevA.103.063711,PhysRevLett.128.073601,PhysRevLett.128.203601,Needham_2019}, population localization and delocalization \cite{PhysRevResearch.6.013159,PhysRevResearch.6.023232,PhysRevResearch.6.013320,Leonard2023}, and enhanced quantum correlations \cite{PhysRevLett.110.080502,PhysRevLett.121.143601,PhysRevA.102.013723,PhysRevLett.126.023603,PhysRevA.105.023717} induced by photon-mediated long-range dipole-dipole interactions \cite{Solano2017}, which generate strongly correlated steady states with potential applications in quantum information processing. This leads to diverse applications, such as photon routing and interference \cite{10.1063/5.0168808}, essential for integrated quantum networks and scalable quantum computation \cite{PCW1,PhysRevA.104.053703}. The extra degree of freedom in controlling coupling directionality \cite{mitsch2014quantum} provides fresh insights into quantum dynamics at these interfaces and paves the way for innovative applications in quantum simulation and quantum computation within next-generation nanophotonic devices.

Recent studies have theoretically employed laser fields incident at Bragg angles to drive the atom-nanofiber system \cite{PhysRevA.104.043517}, with a focus on pronounced resonance of light scattering through a chiral waveguide. The other work has configured a one-dimensional (1D) atomic array in an anti-Bragg periodic layout under strong coherent driving \cite{PhysRevA.106.L031702}. This setting reveals strong subradiant eigenstates featuring long-lived quantum correlations between qubits. In addition, a wide range of intriguing delocalization and localization behaviors has been observed through integrating arrays with a clean and a disordered zone \cite{Leonard2023, PhysRevResearch.6.013159} or with disparate interparticle spacings \cite{PhysRevResearch.6.023232,PhysRevResearch.6.013320}.
\begin{figure*}[t]
	\centering
	\includegraphics[width=17.6cm]{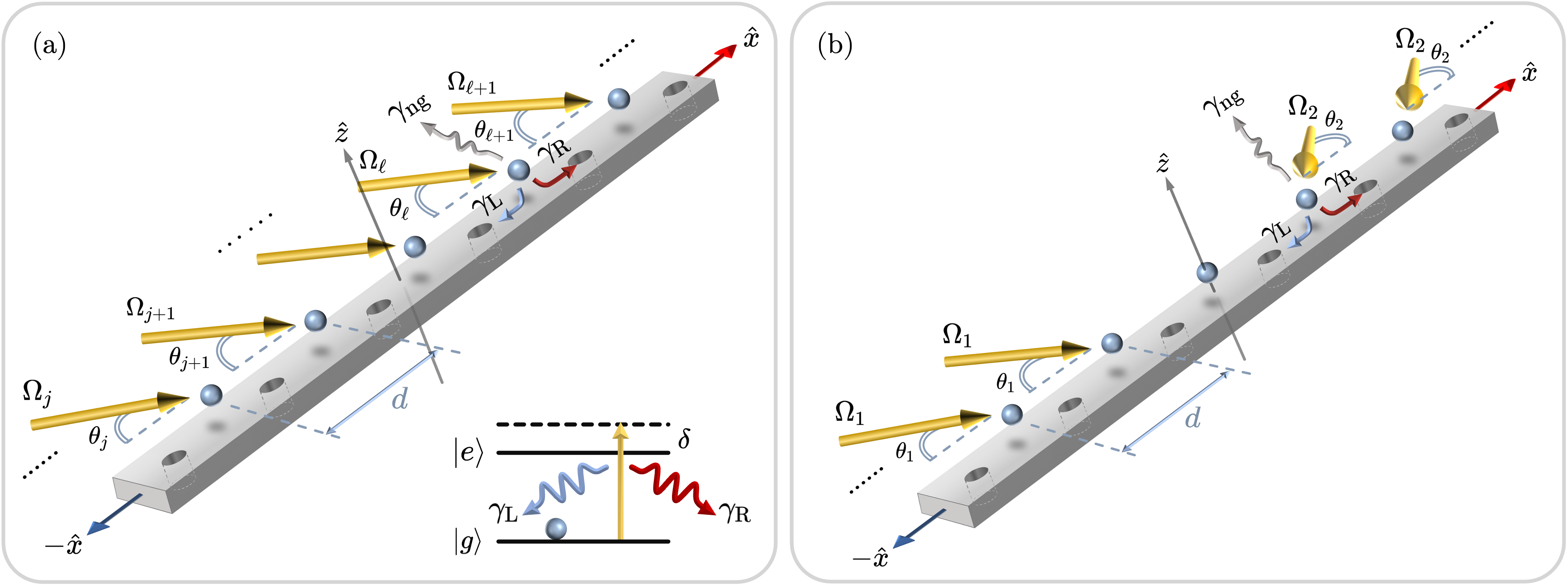}
	\caption{A schematic diagram of a weakly-driven chiral-coupled atomic array. The system consists of a homogeneous one-dimensional array of \(N\) two-level quantum emitters with interparticle distance $d$, coupled to a photonic crystal waveguide. The effective nonreciprocal decay rates \(\gamma_{\text{L}} \neq \gamma_{\text{R}}\) are realized through the guided modes in the waveguide and can be controlled by external magnetic fields, while the non-guided decay \(\gamma_{\text{ng}}\) determines the coupling efficiency. (a) Generalized model: Each atom is weakly driven and excited laterally by laser fields with varying Rabi frequencies \(\{\Omega_{j}\,|\,j = 1,2,\cdots,N\}\), where \(\delta\) represents the detuning from the side excitation. (b) Defect-driving scheme: The driving on one of the atoms is removed, and the left and right chains, centered around this atom, are driven at incident angles \(\theta_1\) and \(\pi - \theta_2\), respectively.}
	\label{fig:schematic plot}
\end{figure*}
Motivated by the aforementioned studies,  here we propose an even straightforward setting in creating a dissimilar driven-dissipative atomic arrays by driving the equidistant atomic array with different incident angles, as shown in Fig.~\ref{fig:schematic plot}.  We anticipate that the phase differences play a vital role in generating interference effects, potentially leading to various intriguing steady states. In certain cases, we observe behaviors reminiscent of those found in a dissimilar array \cite{PhysRevResearch.6.023232}, along with precisely controllable and strong population localization. Notably, such localization effects have also been observed in Rydberg atom systems \cite{Kirova:20}. We investigate various parameters and measures to study strongly confined atomic localization under different excitation schemes, supported by analytical solutions under the reciprocal coupling regime. Our results provide deeper insights into interaction-driven atomic excitations within driven-dissipative systems and open new pathways for the realization of exotic quantum states relevant to quantum information processing.

The structure of this paper is organized as follows: Sec.~\ref{sec:theoretical model} outlines the theoretical framework for a driven-dissipative atomic array with chiral couplings. In Sec.~\ref{sec:Asymmetric Driving Scheme}, we present the steady-state distribution of atomic excitations under asymmetric incident angles and investigate strong interfaced localization arising from different angle combinations. In Sec.~\ref{sec:Defect-driving scheme}, we propose a new scheme with defect driving aimed at enhancing atomic population control and concentration. Finally, Sec.~\ref{sec:conclusion} discusses the implications of our findings and concludes the study.

\section{Theoretical Model}
\label{sec:theoretical model}

This study investigates a chiral coupling interface formed by a driven one-dimensional atomic array and a photonic crystal waveguide in Fig.~\ref{fig:schematic plot}(a). The atomic array comprises $N$ identical two-level atoms (or, more generally, two-level quantum emitters, such as superconducting qubits or quantum dots \cite{RevModPhys.95.015002}) arranged with a uniform interparticle spacing $d$. The photonic crystal waveguide acts as a one-dimensional reservoir \cite{PhysRevLett.110.080502,PhysRevA.91.042116}, transmitting photons in guided modes to mediate the coupling between the atoms through evanescent waves \cite{BLIOKH20151}. Laser fields drive the atoms individually from the side, propagating in a horizontal plane at the same height as the array, which is parallel and close to the $\hat{x}$-$\hat{y}$ plane.

The dynamical evolution of the system's state $\rho$ is governed by the effective chiral Lindblad master equation \cite{PhysRevA.91.042116,PhysRevLett.131.033605} in the interaction picture and under the Born-Markov approximation \cite{PhysRevA.2.883},
\begin{equation}\label{chiral master eq}
	\dv{\rho}{t}
	=
	-\dfrac{i}{\hslash}[H_{\text{S}}+H_{\text{L}}+H_{\text{R}},\rho]
	+
	\mathcal{L}_{\text{L}}[\rho]
	+\mathcal{L}_{\text{R}}[\rho]
	+\mathcal{L}_{\text{ng}}[\rho],
\end{equation}
with Hamiltonians $H_{\text{S}}$ the light-matter interaction from a laser field, $H_{\text{L}(\text{R})}$ the chiral couplings, and Lindblad forms of $\mathcal{L}_{}[\rho]$ indicating the dissipations. The term $H_{\text{S}}$ is
\begin{equation}
	H_{\text{S}}
	=
	\hslash
	\sum_{j=1}^{N}\left[\Omega_{j} e^{ikx_{j}\cos\theta_{j}}\left(
	\sigma_{j}+\sigma_{j}^\dagger\right)
	-\delta_{j}\sigma_{j}^\dagger\sigma_{j}\right],
\end{equation}
which drives $N$ two-level quantum emitters (each characterized by the ground state $\ket{g}_{j}$ and excited state $\ket{e}_{j}$), with  $\Omega_{j}=\tilde{\Omega}_{j}e^{i\phi_{j}}$, where $\tilde{\Omega}_{j}=\Re[\Omega_{j}]$ and $\phi_{j}$ is the corresponding phase for the $j$-th atom.  The emitters are subject to spatially dependent detunings $\delta_{j}$. The dipole operator is defined as $\sigma_{j}^\dagger \equiv \ket{e}_{j}\bra{g}$ with $\sigma_{j} = (\sigma_{j}^\dagger)^\dagger$. The wave number denotes as $k=2\pi/\lambda$ with the wavelength $\lambda$, while the incident angle at $j$-th atom denoted as $\theta_{j}$ characterizes the propagation phases of the laser field. Notably, under normal incidence, where $\cos\theta_{j} = 0$, $H_{\text{S}}$ contains no position-dependent phase terms. With identical $\phi_{j}$, the corresponding phases can be factored out as a global phase, leaving the observable outcomes unaffected. In contrast, under oblique incidence, the position-dependent phase $kx_{j} \cos\theta_{j}$ plays a significant role in both the system's evolution and its steady-state behaviors. Although different choices of the coordinate origin seem to induce distinct system behaviors, these effects can, in principle, be neutralized by appropriately adjusting $\{\phi_{j}\}$. This flexibility in phase tuning enhances the versatility of the platform, enriching the possibilities for system control and application. 

The coherent terms $H_{\text{L}}$ and $H_{\text{R}}$ are
\begin{align}
	H_{\text{L(R)}}
	=
	-i\hslash\dfrac{\gamma_{\text{L(R)}}}{2}
	\sum_{j<(>)\ell}^{N}\left(e^{ik|x_{j}-x_{\ell}|}\sigma_{j}^\dagger\sigma_{\ell}-\text{h.c.}\right),
\end{align}
which represents the collective energy shifts due to the infinite-range photon-mediated dipole-dipole interactions \cite{PhysRevA.95.023838,Solano2017}, and we label the positions of spins such that $x_{j}>x_{\ell}$ when $j>\ell$. The remaining Lindblad terms read
\begin{align}
	\mathcal{L}_{\alpha}[\rho]
	=
	\sum_{j,\ell}^{N}
	\gamma_{j\ell}^{\alpha}
	\left[
	\sigma_{\ell}\rho\sigma_{j}^\dagger
	-\dfrac{1}{2}\{\sigma_{j}^\dagger\sigma_{\ell},\rho\}
	\right],
\end{align}
where $\mathcal{L}_{\text{L}(\text{R})}[\rho]$ represents the collective decay in guided modes with decay rates $\gamma_{j\ell}^{\text{L}(\text{R})}\equiv\gamma_{\text{L}(\text{R})}e^{+(-)ik(x_{j}-x_{\ell})}$, using the subscripts L (R) to label the decay channels that propagate to the left (right). While $\mathcal{L}_{\text{ng}}[\rho]$ corresponds to the non-guided decay experienced by the atoms, with decay rate $\gamma_{j\ell}^{\text{ng}}\equiv \gamma_{\text{ng}}\delta_{j\ell}$, which is intrinsic and is considered to be the same for all identical atoms. For instance, in an atom-nanofiber system, approximately $99\%$ of light is typically scattered due to free-space decay \cite{RevModPhys.95.015002}. In contrast, in systems where quantum dots or superconducting qubits are coupled to a photonic crystal waveguide, free-space scattering can be reduced to around $0.5\%$ on average \cite{RevModPhys.95.015002,PCW1,PhysRevLett.113.093603}, yielding significantly enhanced coupling efficiency. To quantify the tendency of chiral coupling arising from the competition between $\gamma_{\text{R}}$ and $\gamma_{\text{L}}$, the directionality of the couplings $D \equiv (\gamma_{\text{R}}-\gamma_{\text{L}})/\gamma$ \cite{mitsch2014quantum} is introduced. Here, the total guided decay rate can be written as $\gamma\equiv\gamma_{\text{R}} + \gamma_{\text{L}} \equiv 2|\dd q(\omega)/\dd \omega|_{\omega =\omega_{\text{eg}}}g_{k}^{2}L$ \cite{PhysRevLett.110.080502}, where $|\dd q(\omega)/\dd\omega|_{\omega = \omega_{\text{eg}}}$ represents the inverse of group velocity at resonance, with the atom-waveguide coupling strength $g_k$ and the quantization length $L$.

We initialize the system in the ground state $\ket{g}^{\otimes N}$, and consider the system dynamics under weak excitation, namely $\Omega_{j}\ll \gamma_{j\ell}^{\alpha}$ \cite{PhysRevA.104.043517}. This assumption confines dynamical evolution to a self-consistent Hilbert subspace $\{\ket{g}^{\otimes N}, \ket{\psi_{j}}=\ket{e}_{\mu}\ket{g}^{\otimes (N-1)}\}$ for $j\in[1,N]$, restricted to the ground state and the manifold of singly excited states. Thus, the total state can be written as
\begin{equation}
	\ket{\Psi(t)}=\dfrac{1}{\sqrt{1+\sum_{j=1}^{N}|p_{j}(t)|^2}}\left[\ket{g}^{\otimes N} +\sum_{j=1}^{N}p_{j}(t)\ket{\psi_{j}}\right],
\end{equation}
where $p_{j}(t)$ represents the probability amplitude and $\sum_{j=1}^{N}|p_{j}(t)|^2\ll 1$ to satisfy the weak-excitation assumption. Thus Eq.~(\ref{chiral master eq}) can be reduced to the coupled equations for $p_{j}(t)$ \cite{PhysRevResearch.2.013097} as
\begin{equation}
	\dot{p}_{j}=-i\Omega_{j}e^{ik\phi_{j}} e^{ikx_{j}\cos\theta_{j}}+\sum_{j=1}^{N}[\vb{M}]_{j\ell}p_{j},
	\label{probability amplitude}
\end{equation}
where the matrix elements of the coupling matrix $\vb{M}$ are:
\begin{equation}\label{coupling matrix}
	[\vb{M}]_{j\ell} =
	\begin{cases}
		-\gamma_{\text{L}} e^{ik|x_{j}-x_{\ell}|}&,\; j < \ell
		\\
		i\delta_{j}-\frac{\gamma_{\text{L}}+\gamma_{\text{R}}+\gamma_{\text{ng}}}{2}&,\; j = \ell
		\\
		-\gamma_{\text{R}} e^{ik|x_{j}-x_{\ell}|}&,\; j > \ell
	\end{cases}.
\end{equation}
We then define the dimensionless interparticle distance $\xi\equiv k(x_{j}-x_{j-1})$. Consequently, the steady-state probability amplitudes, satisfying $\dot{p}_{j}=0$, are given by
\begin{equation}
	\tilde{p}_{j}\equiv p_{j}(t\to\infty)
	=i \sum_{\ell=1}^{N}[\vb{M}^{-1}]_{j\ell}\Omega_{\ell}e^{i\phi_{\ell}}e^{ikx_{\ell}\cos\theta}.
	\label{probability amplitude of steady state}
\end{equation}
 
From Eqs.~(\ref{probability amplitude}), (\ref{coupling matrix}), and (\ref{probability amplitude of steady state}), we can identify interaction-driven quantum behaviors of atomic excitations, primarily governed by the directionality $D$, photon-mediated dipole-dipole interactions quantified by $\xi$, and the incident angular configurations $\{\theta_{j}\}$. In the following sections, we proceed to characterize the localized steady states that emerge within a homogeneous atomic array under different driving schemes, and we further investigate the strongly confined atomic localization achievable within specific parameter regimes.

\section{Asymmetric Driving Scheme}
\label{sec:Asymmetric Driving Scheme}
\subsection{Steady-State Distribution}
In this section, we obtained the steady-state distribution for a 1D homogeneous array consisting of $N$ atoms under the asymmetric driving scheme at resonance, i.e., $\delta_{j}=0$. We denote the incident angles as
\begin{equation}
	\theta_{j}=
	\begin{cases}
		\theta_{1}&,\,j\in[1,m]
		\\
		\theta_{2}&,\,j\in[m+1,N]
	\end{cases},
	\label{asymmetric driving scheme}
\end{equation}
where $m=\lceil N/2\rceil$. Additionally, we consider the Rabi frequencies of the laser fields from both sides to be $\tilde{\Omega}_{1}=\tilde{\Omega}_{2}$ and set $\phi_{1(2)} = -k_{0} x_{1} \cos \theta_{1(2)}$. This setup uses the first atom's position as the coordinate reference point. Under these conditions, we performed numerical calculations for the normalized steady-state population distribution $\tilde{P}_{j}$ defined as
\begin{equation}
	\tilde{P}_{j}\equiv\dfrac{|\tilde{p}_{j}|^2}{\sum_{j=1}^{N}|\tilde{p}_{j}|^2}.
	\label{normalized population}
\end{equation}

Concerning the simplest case of a homogeneous array subjected to uniform normal incidence $\theta_{1} =\theta_{2}= \pi/2$, we obtained the steady-state phase diagram in the parameter space $(D, \xi)$ \cite{PhysRevResearch.2.013097} using the inverse participation ratio (IPR) \cite{PhysRevB.83.184206}, defined as
\begin{equation}
	\text{IPR}
	\equiv
	\dfrac{\sum_{j=1}^{N}\left(\Delta \tilde{P}_{j}\right)^2}{\left(\sum_{j=1}^{N}\Delta\tilde{P}_{j}\right)^2},
	\label{IPR}
\end{equation}
where $\Delta \tilde{P}_{j}=|\tilde{P}_{j}-N^{-1}|\Theta(\tilde{P}_{j}-N^{-1})$ with the Heaviside step function $\Theta$. It indicates that one of the $\Delta \tilde{P}_{j}$ dominates as  $\text{IPR} \to 1$, signifying strong localization. Conversely, delocalized behaviors are recognized as $\text{IPR} \to N^{-1}$. We classified five distinguishable steady-state phases through the phase diagram \cite{PhysRevResearch.2.013097}, which are
\begin{enumerate}[(1)]
	\itemsep = -0.07cm
	\item uniformly extended distributions (ETD) when  $\xi\approx 0$;
	\item the crystalline ordered (CO) phase possessing an finite structure factor, mostly for a finite $D$;
	\item the bi-edge (BE) excitations;
	\item the bi-hole excitations (BH) with hole excitations at the edges mostly for low $D$; and
	\item the chiral-flow dichotomy (CFD) when $\xi=\pi$, which depends on the parity of $N$: a linear slope for an even $N$ and a concave curve for an odd $N$.
\end{enumerate}

Among these, the ETD, CO, and BH phases exhibit delocalized characteristics, while the BE phase demonstrates localized properties. Furthermore, two critical parameters of reciprocal coupling ($D=0$) with $\xi=\{0\,(2\pi),\pi\}$ can be identified and excluded from the steady-state phases. This exclusion is due to constructive interference in state populations — a characteristic of decoherence-free space \cite{PhysRevLett.81.2594} being predominantly populated. This leads to a breakdown of the weak-excitation assumption. In the case of dissimilar arrays formed by two segments with different interparticle spacings \cite{PhysRevResearch.6.023232,PhysRevResearch.6.013320}, the steady-state population distribution is mostly determined by a combination of the five aforementioned steady states, as expected. However, a few combinations exhibit new localized patterns such as half-depletion (HD) phases due to the presence of the interface atom. This highlights the complexity and the effect of interfaces within interacting quantum systems.
\begin{figure}[htb]
	\centering
	\includegraphics[width=8.5cm]{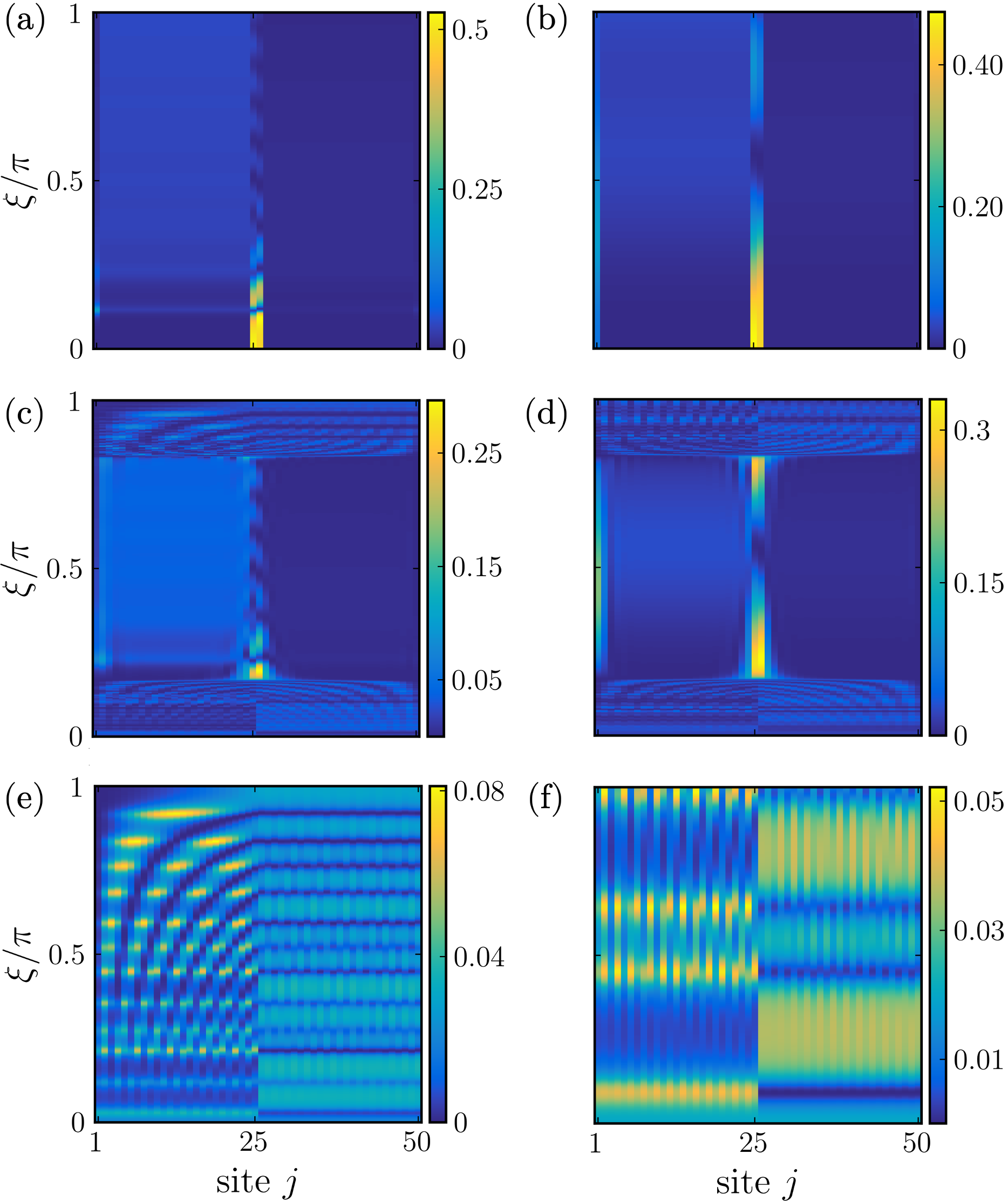}
	\caption{The steady-state population distributions $\tilde{P}{j}$ for $N=50$ atoms, subject to the asymmetric driving scheme. The distributions are presented as a function of $\xi$, with varying angular configurations $(\theta_{1}, \theta_{2})$ and directionality $D$. Panels (a), (c), and (e) correspond to $(\theta_{1}=\pi/2,\theta_{2}=\pi/4)$, while (b), (d), and (f) represent $(\theta_{1}=\pi/4,\theta_{2}=\pi/6)$. The directionality $D$ is set to 0, 0.5, and 1 for [(a), (b)], [(c), (d)], and [(e), (f)], respectively, as comparisons.}
	\label{fig:figure2}
\end{figure}
\begin{figure*}[t]
	\centering
	\includegraphics[width=17.6cm]{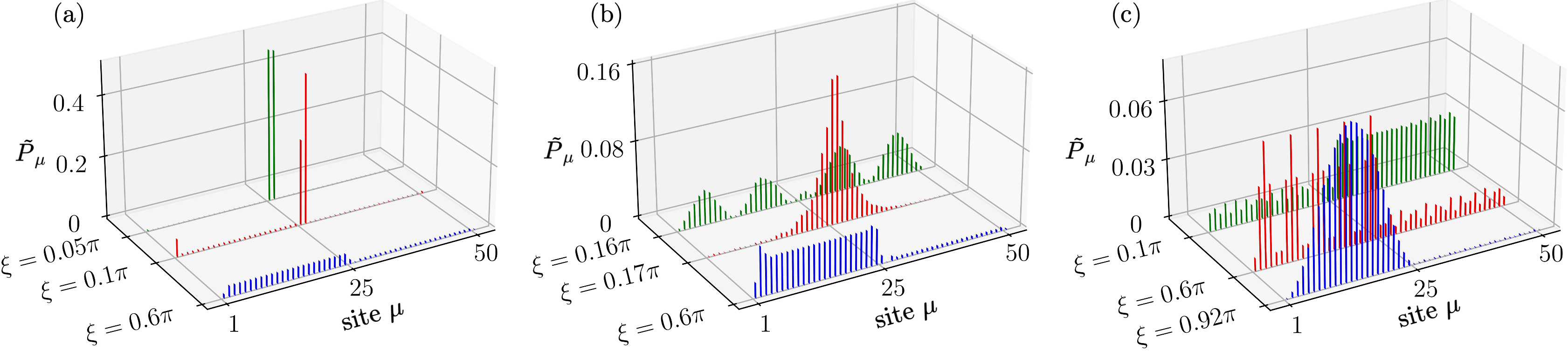}
	\caption{
		Some examples of steady-state population distributions $\tilde{P}_{\mu}$. Panels (a), (b), and (c) correspond to specific cross-sections from Fig.~\ref{fig:figure2}(a), (c), and (e), respectively. In panel (a), for some certain values of $\xi$ (e.g., $0.05\pi$ and $0.1\pi$), strongly confined localization at the interface atoms can be observed. Other confined localizations, though not as extreme as in (a), can also be observed in panels (b) and (c). These soliton-like distributions appear at the center or in the other half of the array.}
	\label{fig:figure3}
\end{figure*}
In Fig.~\ref{fig:figure2}, we demonstrate the steady-state population distribution $\{\tilde{P}_{j}\}$ for $N=50$ in terms of parameter space $(\xi, D, \theta_{1}, \theta_{2})$,
where some of the cross-section plots are illustrated in Fig.~\ref{fig:figure3}. In Figs.~\ref{fig:figure2}(e) and (f), we observe a clear boundary in the steady-state population between the $m$-th and $(m+1)$-th atoms, leading to a steady-state distribution resembling that in a dissimilar array \cite{PhysRevResearch.6.023232}. This phenomenon is more clearly illustrated in cross-sections of Fig.~\ref{fig:figure2}(e), as desplayed in Fig.~\ref{fig:figure3}(c). Under asymmetric driving, unexpected steady-state distributions emerge, such as CO-CO configurations and even a HD-like combination of a soliton-like and a depleted state. These distributions showcase that most of the atomic populations tend to distribute in one of the segments under unidirectional coupling. By contrast, Figs.~\ref{fig:figure2}(a), \ref{fig:figure2}(b), \ref{fig:figure2}(d) showcase a widespread range of interfaced localization, particularly in the lower values of $D$ and $\xi$. Specifically, the green bar shown in Fig.~\ref{fig:figure3}(a), corresponding to
Fig.~\ref{fig:figure2}(a) at $\xi=0.05\pi$, indicates that the population is strongly confined at the $m$-th and $(m+1)$-th atoms. Essentially, the two-site localization can be further manipulated to shift to different locations by simply adjusting the interface position via external laser drivings, allowing for precise control over the localization. As a comparison, the system reaches a steady state of bi-edge excitation (with $\tilde{P}_{1,N} \approx 0.314$ and $\tilde{P}_{j \neq 1,N} \approx 0.0078$) when driven by a uniform normal incidence under the same parameters $\{D, \xi\}$ of green bars in Fig.~\ref{fig:figure3}(a). Interestingly, the localized population can be shifted from the edges and concentrated on the two interface atoms simply by applying external fields at asymmetric angles.

\subsection{Strongly Confined Localization}

\begin{figure}[htp]
	\centering
	\includegraphics[width=8.5cm]{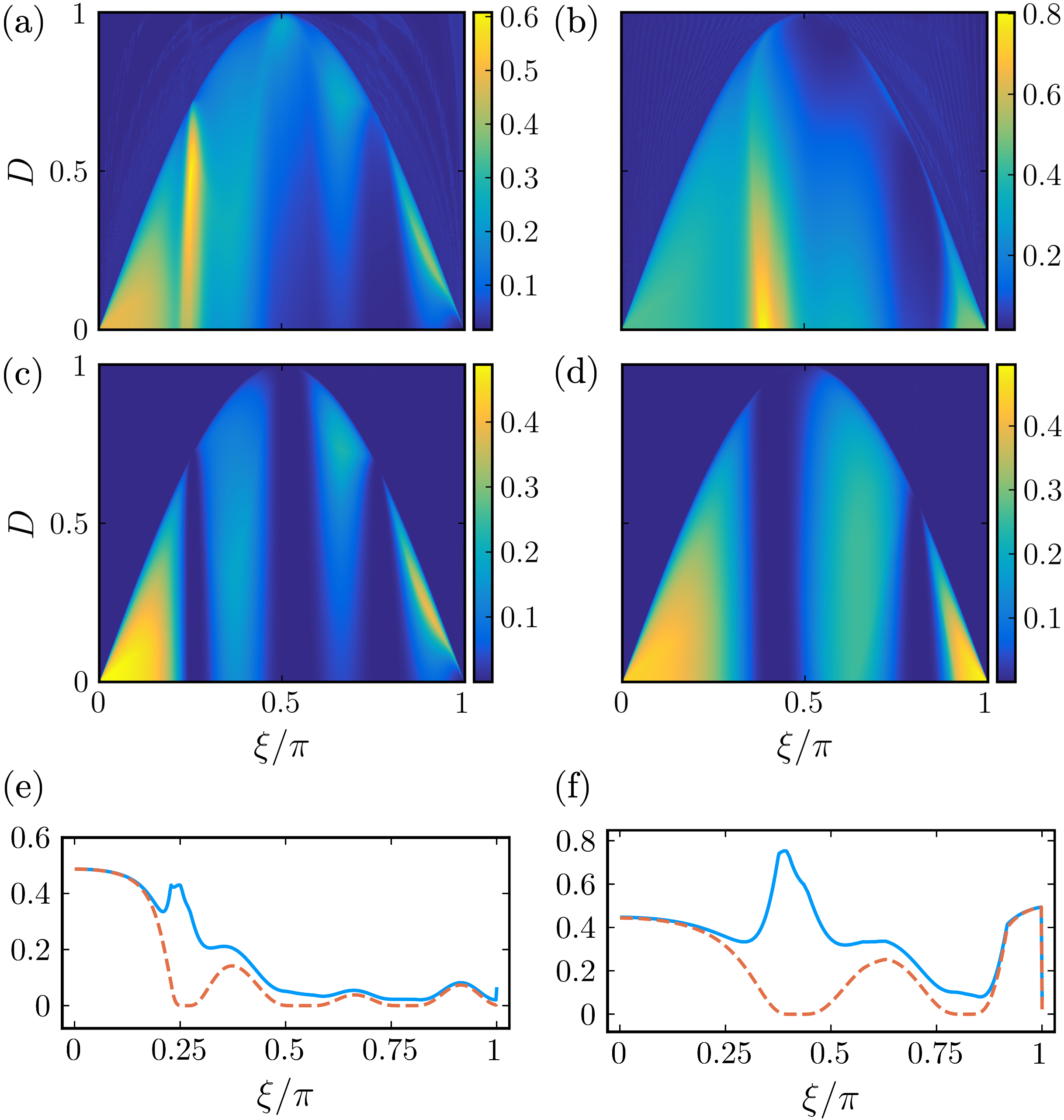}
	\caption{In the asymmetry-driven scheme, the inverse participation ratio (IPR) and interface inverse participation ratio (IIPR) distributions are presented for $N=100$ atoms. Panels (a) and (c) show the IPR and IIPR respectively at $(\theta_{1} = \pi/4, \theta_{2} = \pi/6)$, while (b) and (d) correspond to $(\theta_{1} = \pi/6, \theta_{2} = \pi/12)$.  Cross-sections at $D=0$ from [(a), (c)] and [(b), (d)] are depicted in (e) and (f), respectively, comparing the differences between the IPR (blue solid line) and IIPR (orange dashed line).}
	\label{fig:figure4}
\end{figure}

Next, we seek to delineate the parameter regions where localization emerges. This can be achieved using the IPR defined in Eq.~(\ref{IPR}), with the corresponding simulation results presented in Figs.~\ref{fig:figure4}(a) and \ref{fig:figure4}(b). Within the arched region, several highly intense elongated zones, as well as a triangular area in the lower-left corner, are evident. Nevertheless, these high-intensity regions merely signify the existence of strongly confined localization and do not  exclusively indicate the interfaced localization. In fact, strong localization in this system is not restricted to the atoms around the interface. A closer examination of Fig.~\ref{fig:figure2} reveals that, in certain scenarios, the population becomes localized at one of the terminal atoms of the array, resulting in a single-edge excitation. For example, such behavior can be observed in the cross-section at $\xi \approx 0.6\pi$ in Fig.~\ref{fig:figure2}(b). To rigorously distinguish interfaced localization from single-edge excitation, we define the following measure called the interfaced IPR (IIPR):
\begin{equation}
	\text{IIPR}\equiv
	\dfrac{\sum_{j'\in\text{interface}}(\Delta P_{j'})^{2}}{\left(\sum_{j=1}^{N}\Delta P_{j}\right)^2},
	\label{eq: IIPR}
\end{equation}
where the interface atoms are located at $j' \in \{m, m+1\}$ under the asymmetric driving scheme. If the IPR attains a maximum for certain parameters and coincides with the IIPR, this signifies interfaced localization. Conversely, if the two measures do not overlap and the IIPR reaches a minimum, it indicates the presence of single-edge excitation.

In Figs.~\ref{fig:figure4}(c) and \ref{fig:figure4}(d), we present corresponding IIPR distributions of the parameters from Figs.~\ref{fig:figure4}(a) and \ref{fig:figure4}(b), respectively. The results indicate that bright elongated regions in the IPR distribution disappear in the IIPR map, while the remaining bright areas primarily are concentrated in the triangular region at the lower left corner, which exactly corresponds to where interfaced localization occurs. This result further suggests that interfaced localization is most likely observed when the value of $D$ is small, which aligned with the phenomenon we observed in Fig.~\ref{fig:figure2}. One can distinguish between these two types of excitation localization more clearly by comparing the cross-sections of the IPR and IIPR at $D=0$, illustrated in Figs.~\ref{fig:figure4}(e) and \ref{fig:figure4}(f), respectively. For instance, in Fig.~\ref{fig:figure4}(e), interfaced localization manifests in the range $\xi \in (0, 0.15\pi]$, while one-edge excitation emerges at $\xi = 0.25\pi$. Likewise, in Fig.~\ref{fig:figure4}(f), the interfaced localization can be revealed in two distinct intervals: $\xi \in (0, 0.2\pi]$ and $\xi \in [0.9\pi, \pi)$. In contrast, single-edge excitation appears around $\xi = 0.4\pi$. It should be noted that our analysis excludes scenarios involving extremely small interparticle separations, where near-field effects become significant \cite{PhysRevA.98.013849,PhysRevA.101.053852}. Thus, for an interatomic separation value of around $\xi \geq 0.1\pi$, these results could still remain valid. Overall, our defined IIPR and IPR calculations can faithfully identify the parameter regions that can distinguish the interfaced localization from the single-edge excitation localization.

\subsection{Size Effect on Interfaced Localization}
\begin{figure}[htb]
	\centering
	\includegraphics[width=8.2cm]{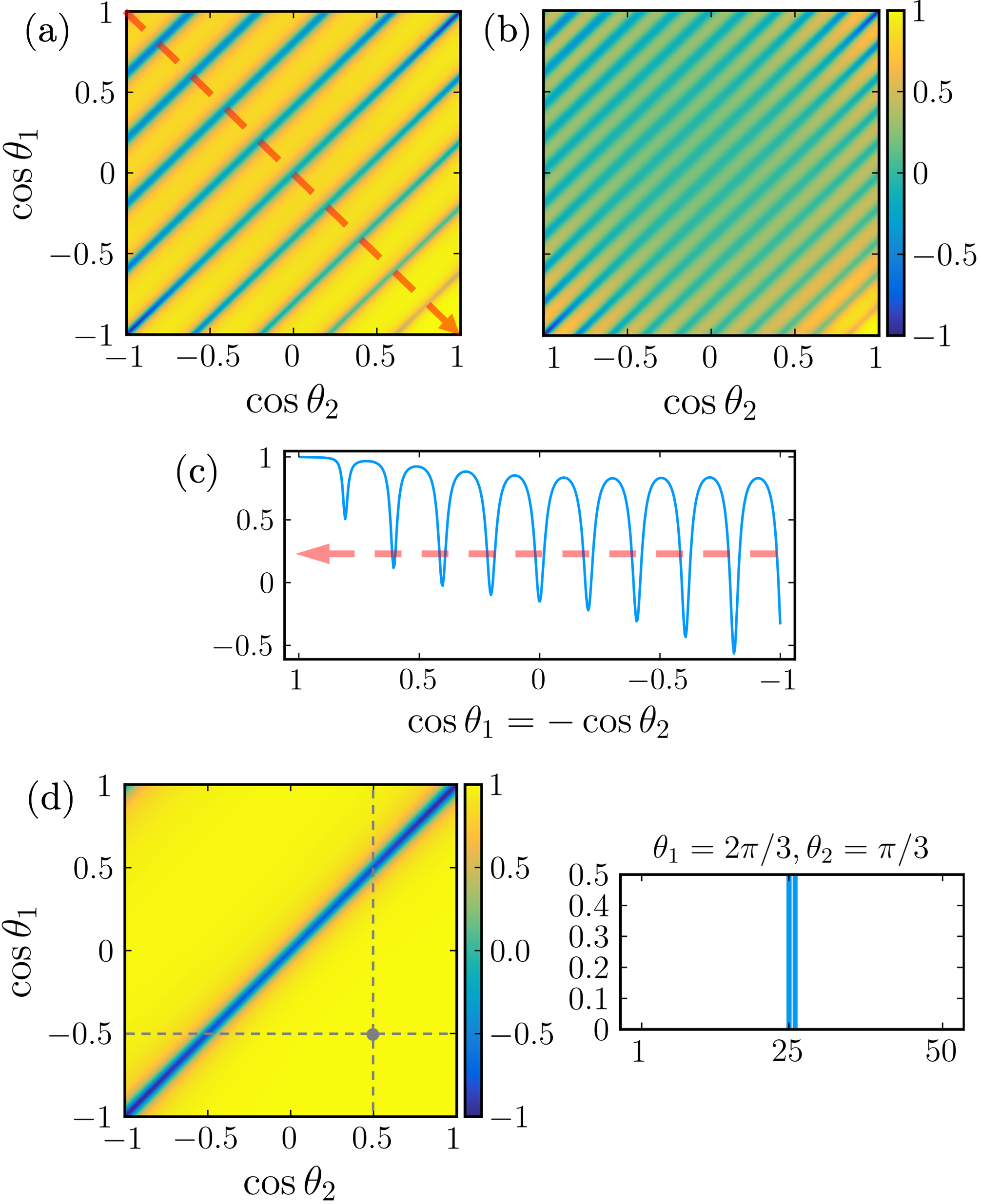}
	\caption{In the asymmetry-driving scheme, the RIEL is presented with respect to $(\cos\theta_{1}, \cos\theta_{2})$. Panel (a) shows the RIEL distribution for $\xi=0.1\pi$, while (b)corresponds to the case of $\xi=0.2\pi$, both for $N=100$. A cross-section along the diagonal of (a) is illustrated in (c), showcasing the distribution of RIEL's local minima. The left subplot in panel (d) shows the RIEL distribution for $N=50$ and $\xi=0.4\pi$. The right subplot displays the double-site atomic localization at the interface corresponding to the gray dot parameters $\theta_{1}=2\pi/3$ and $\theta_{2}=\pi/3$ in the RIEL distribution.}
	\label{fig:figure5}
\end{figure}

Here, we examine the excitation localizations under varying $(\theta_{1}, \theta_{2})$ with fixed $(D, \xi)$. To further distinguish between interfaced localization and edge excitation, we introduce the ratio of interface-edge localization (RIEL), defined as
\begin{equation}
	\text{RIEL}=\dfrac{\tilde{P}_{\text{interface}}-\tilde{P}_{\text{edge}}}{\sum_{j=1}^{N}\tilde{P}_{j}},
	\label{eq: RIEL}
\end{equation}
where $\tilde{P}_{\text{interface}}=\tilde{P}_{m}+\tilde{P}_{m+1}$ and $\tilde{P}_{\text{edge}}=\tilde{P}_{1}+\tilde{P}_{N}$. As $\text{RIEL}\to 1$ ($-1$), the population distribution approaches and resembles interfaced localization (edge excitation). Here, we focus on the RIEL distribution under reciprocal coupling conditions ($D=0$), where interfaced localization is most likely to manifest in low-$D$ regimes, as shown in Fig.~\ref{fig:figure5}. Figures~\ref{fig:figure5}(a) and \ref{fig:figure5}(b) reveal that RIEL exhibits a mirror-symmetric structure with respect to the diagonal ($\cos\theta_{1}=-\cos\theta_{2}$) in the $(\cos\theta_{1}, \cos\theta_{2})$ space, forming banded structures. The triangle area in the lower right corner and other prominent bands suggest a higher propensity for interfaced localization when driven by the combination with a large $\theta_{1}$ and a small $\theta_{2}$. For instance, in Fig.~\ref{fig:figure5}(a), when $(\theta_{1}, \theta_{2}) = (0.9\pi, 0.1\pi)$, $\tilde{P}_{\text{interface}} \approx 0.998$ is observed and similarly for $(\theta_{1}, \theta_{2}) = (0.75\pi, 0.25\pi)$, we find $\tilde{P}_{\text{interface}} \approx 0.968$ 

Moreover, these layered band structures exhibit angular periodicity associated with $\xi$. For a fixed system size $N$, five prominent bands appear within the lower triangular region of the heatmap when $\xi = 0.1\pi$ in Fig.~\ref{fig:figure5}(a). As $\xi$ doubles, this region manifests ten bands in Fig.~\ref{fig:figure5}(b). With further increases in $\xi$, these bands gradually compress and merge, converging toward zero values, indicating delocalized distributions. Meanwhile, the lower-right triangle region retains its high intensity, despite steady reduction in size. Interfaced localization can only be sustained under asymmetrical driving of a larger $\theta_1$ paired with a smaller $\theta_2$. Such configurations effectively counteract the system's natural delocalization tendency induced by greater atomic separation.

Furthermore, by analyzing cross-sections along the diagonal (where $\cos\theta_{1} = -\cos\theta_{2}$) of this distribution, we establish an empirical formula that captures the local minima of RIEL, which is
\begin{equation}
	\cos\theta_{1}-\cos\theta_{2}=\dfrac{2n\pi}{m\xi}
	\;,\;\; n\in\mathbb{Z}
	\;,\;\;
	\xi\in(0,\pi),
	\label{eq: size effect}
\end{equation}
where $n\in[1-(m\xi/\pi),(m\xi/\pi)-1]$, and $m=\lceil N/2\rceil$ related to the system size $N$. For example, in Figs.~\ref{fig:figure5}(a) and \ref{fig:figure5}(c), where $m\xi = 5\pi$, each local minimum along the diagonal is observed at $\cos\theta_{1} = -\cos\theta_{2} = n/5$, where $n \in[-4,4]$. Therefore, Eq.~(\ref{eq: size effect}) facilitates the prediction of RIEL distributions under $D=0$ for different $(N, \xi)$. Notably, as $m\xi \to \pi$, the RIEL displays a single trench of local minima along the anti-diagonal (where $\cos\theta_{1} = \cos\theta_{2}$), with high-intensity regions predominating elsewhere. The results presented in Fig.~\ref{fig:figure5}(d) show excellent agreement with the prediction by Eq.~(\ref{eq: size effect}). This indicates that the majority of angular combinations are inclined toward interfaced localization, apart from cases where $\theta_{1} = \theta_{2}$. Conversely, when $m\xi \gg \pi$, concentrating atomic populations at the interface becomes progressively difficult, as more minimums emerge based on Eq.~(\ref{eq: size effect}). These patterns collectively underscore the substantial influence of system size on the formation of interfaced localization.

\section{Defect-driving scheme}
\label{sec:Defect-driving scheme}
In the previously mentioned scheme, the phase distribution of the driving fields experienced by each atom is asymmetric. To explore more intriguing excitation localization in the steady states, we introduce the defect-driving scheme, where the driving field can be removed from one or multiple atoms, as illustrated in Fig.~\ref{fig:schematic plot}(b). In this scheme, it is assumed that the interatomic spacing is sufficiently large such that the fields $\Omega_{1}$ and $\Omega_{2}$ do not unintentionally excite the $m$-th atom. The incident angles for the remaining atoms are defined as
\begin{equation}
	\theta_{j}=\begin{cases}
		\theta_{1}\;&,\,j\in[1,m-1]
		\\
		\pi-\theta_{2}\;&,\,j\in[m+1,N]
	\end{cases}.
	\label{eq: scheme2}
\end{equation}
The undriven $m$-th atom distinguishes the array into a left- and a right chain, for which the phase reference point is set at the $(m-1)$-th  and the $(m+1)$-th atom, respectively. This symmetry in phase distribution can be achieved by defining the Rabi frequencies with phases $\phi_{1} = -k_{0}x_{m-1}\cos\theta_{1}$ and $\phi_{2} = -k_{0}x_{m+1}\cos\theta_{2}$. Thus, the phase profile of the system attains symmetry with respect to the undriven atom. In this section, we focus on the defect-driving scheme under $D=0$, as we believe and will demonstrate that, akin to asymmetrical driving schemes, this setup will yield pronounced interfaced localizations under reciprocal coupling.
\begin{figure}[htb]
	\centering
	\includegraphics[width=8.5cm]{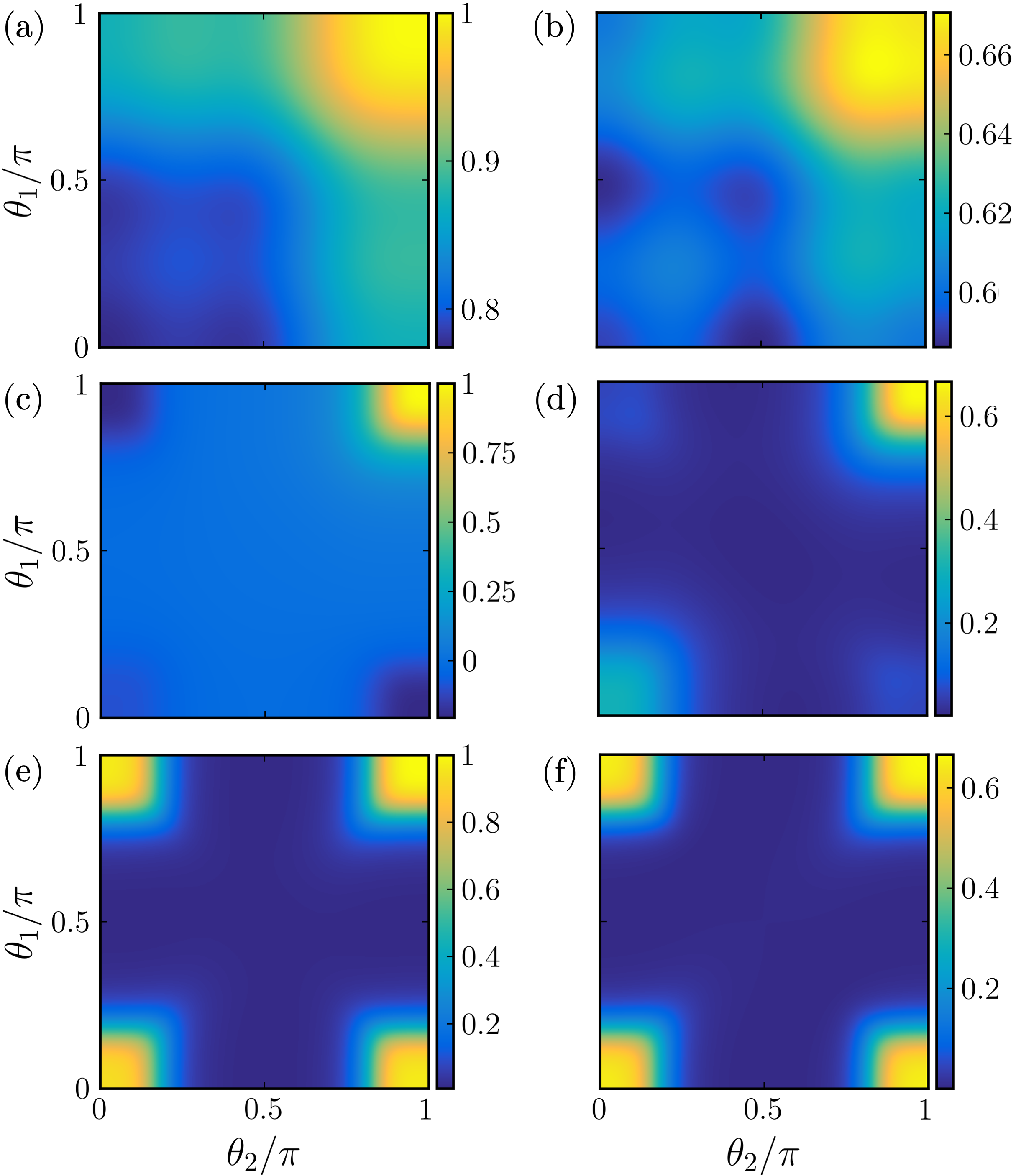}
	\caption{Within the defect-driven scheme, the distributions of the RIEL (panels (a), (c), and (e)) and  the population of the non-driven atom $\tilde{P}_{m}$ (panels (b), (d), and (f)) are presented in the space $(\theta_{1}, \theta_{2})$. Panels [(a), (b)], [(c), (d)], and [(e), (f)] correspond to $\xi=0.1\pi$, $\xi=0.4\pi$, and $\xi=0.95\pi$, respectively, showcasing the variation in distributions as the interatomic spacing increases. Other parameters are chosen at $N=100$ and $D=0$.}
	\label{fig:figure6}
\end{figure}
\subsection{Saturated Population in Interfaced Localization}

We first calculated the RIEL distribution for this driving scheme, where $\tilde{P}_{\text{interface}} = \tilde{P}_{m-1} + \tilde{P}_{m} + \tilde{P}_{m+1}$, with results shown in Figs.~\ref{fig:figure6}(a), \ref{fig:figure6}(c), and \ref{fig:figure6}(e). Notably, in most parameter regimes, RIEL remains non-negative, indicating the absence of strongly edged excitations. However, strong interfaced localizations are still observable. We can therefore identify the parameter ranges where strongly confined atomic localization occurs by examining only the population at the interface atoms.

\begin{figure*}[t]
	\centering
	\includegraphics[width=16cm]{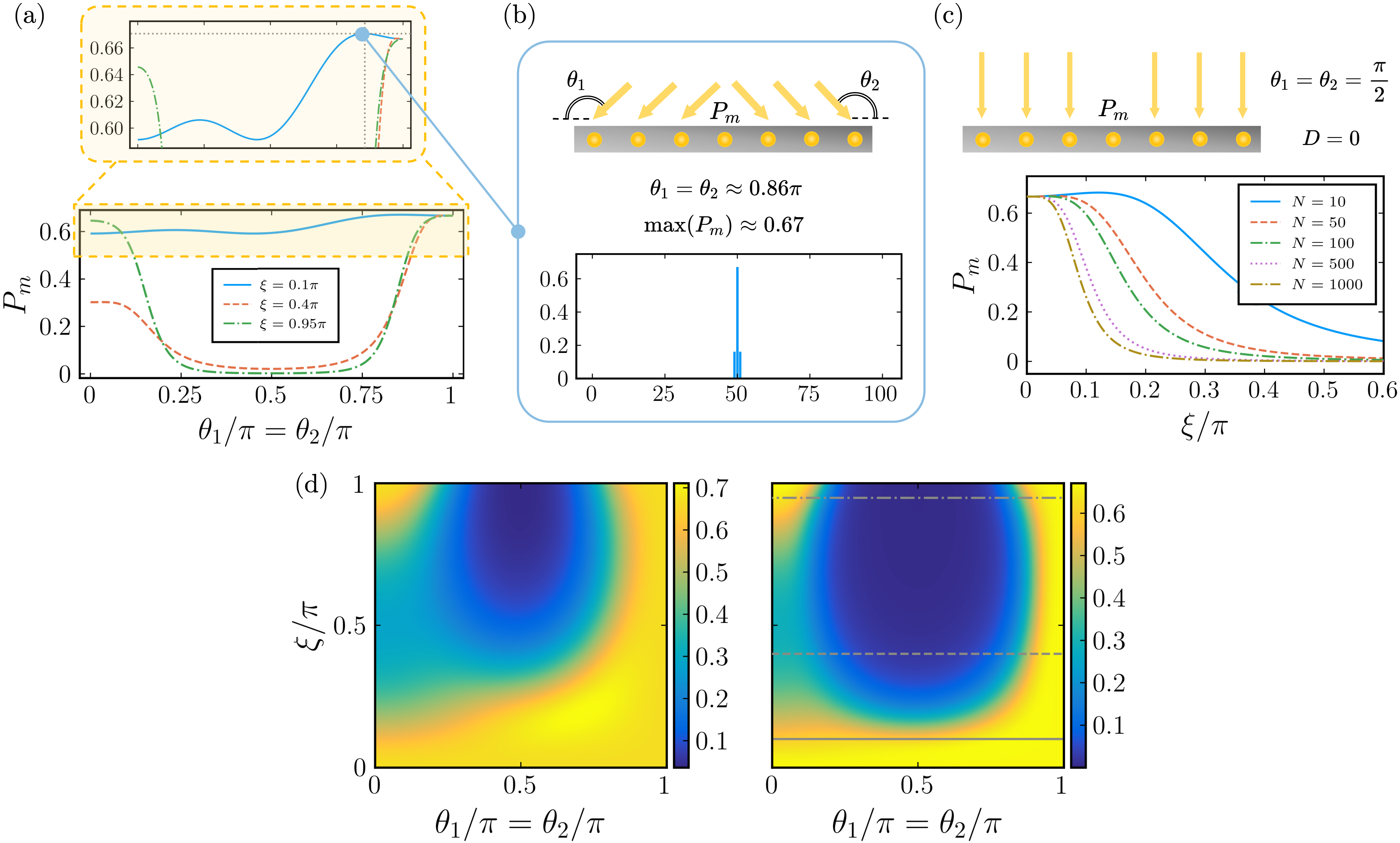}
	\caption{The saturation of the non-driven atom population $\tilde{P}_{m}$. (a) presents cross-sections along the anti-diagonal from Figs. 6(b), 6(d), and 6(f), depicting the variation of $P_{m}$ with $\theta_{1} = \theta_{2}=\theta$ for $D=0$ and specific interatomic distances: $\xi=0.1\pi$ (blue solid line), $\xi=0.4\pi$ (orange dashed line), and $\xi=0.95\pi$ (green dotted line). (b) showcases the corresponding maximum of the blue solid line ($D=0$, $\xi=0.1\pi$) from (a), showing that under antisymmetric driving at $\theta_{1} = \theta_{2} \approx 0.86\pi$, $\tilde{P}_{m}$ reaches a maximum value of approximately $0.67$. (c) demonstrates how $\tilde{P}_{m}$ varies with interparticle distance $\xi$ under normal incidence for different atom numbers $N=10$ (blue solid line), $N=50$ (orange dashed line), $N=100$ (green dash-dot line), $N=500$ (purple dotted line), and $N=1000$ (brown dash-dot line). The results reveal that as $\xi \to 0$ (i.e., $2\pi$), $\tilde{P}_{m}$ saturates to exactly $2/3$, regardless of $N$. (d) compares the $\tilde{P}_{m}$ distribution for $N=10$ and $N=100$ in the left and right subplots, respectively, where the gray solid-, dashed-, and dash-dotted line correspond to the curves with the same styles in (a).}
	\label{fig:figure7}
\end{figure*}

Next, we analyze the distribution of $\tilde{P}_{m}$, the population at the undriven atom across various anglular combinations $(\theta_{1}, \theta_{2})$ for specific values of $\xi$, as shown in Figs.~\ref{fig:figure6}(b), \ref{fig:figure6}(d), and \ref{fig:figure6}(f). For $\xi<0.5\pi$, the maximum values of $\tilde{P}_{m}$ appear concentrated in a highlighted region in the upper right corner. As $\xi$ approaches $\pi$, high-intensity regions emerge in all four corners of the heatmap, creating a perfectly symmetric distribution. Subsequently, we examine cross-sections along the anti-diagonal (where $\theta_{1} = \theta_{2} = \theta$) to compare the maximum values of $\tilde{P}_{m}$ across three distinct $\xi$ values, as depicted in Fig.~\ref{fig:figure7}(a). These results suggest that a considerable population proportion on the undriven atom can most likely be observed under a small value of $\xi$, e.g., the profile for $\xi = 0.1\pi$ mostly surpasses those of the other cases significantly, reaching a peak of $\tilde{P}_{m} \approx 0.67$ at $\theta = 0.86\pi$. The corresponding driving configuration and resulting steady-state population distribution for this scenario are demonstrated in Fig.~\ref{fig:figure7}(b), revealing the strongest single-site excitation observed at the interface in this study so far, along with two accompanying tiny side lobes at the $(m-1)$-th and the $(m+1)$-th atoms. In fact, this localization demonstrates robustness against the non-guided decay up to $\gamma_{\text{ng}}\approx 0.5\gamma$.

To further investigate the effect of $\xi$ on $\tilde{P}_m$, we performed a scan over the range $\xi \in (0, \pi)$, with the findings displayed in Fig.~\ref{fig:figure7}(d). The regions, where the strong interfaced localization is most likely to occur, are primarily located along the right edge and bottom of the plot, covering a combined region of the entire $\theta$ range at low $\xi$ values and the full $\xi$ range at high $\theta$ values. Additionally, one can observe dark regions shaped like cave entrances where $\tilde{P}_{m}$ approaches 0, forming a hole excitation at the undriven atom. This dark region expands as the system size $N$ increases, effectively compressing the parameter area for generating interfaced localization. To further corroborate this phenomena, we examine the vertical cross-section of these colormaps under $\theta_{1} = \theta_{2} = \pi/2$ for various system sizes, with $N$ ranging from 10 to 1000, as shown in Fig.~\ref{fig:figure7}(c). It is clear that as the system size grows, the maximally attainable $\tilde{P}_{m}$ converges to a numerical value of approximately $0.667$, which we refer to as the “saturated population." Concurrently, the interparticle distance $\xi$, where saturation occurs, approaches 0 (or $2\pi$). Such behaviors can be explained through the analytical solution presented below in Eq.~(\ref{analytical solution}).
\subsection{Analytical Solutions}
Here, we present the analytical solutions for the configuration depicted in Fig.~\ref{fig:figure7}(c), where $\theta_{1} = \theta_{2} = \pi/2$ and $D = 0$. The steady-state probability amplitudes $\{\tilde{p}_{j}\}$ can be derived analytically through Eq.~(\ref{probability amplitude of steady state}) and are given by
\begin{equation}
	\tilde{p}_{j} =-\Omega \begin{cases}
		i+\tan\left(\xi/2\right) & ,\;j=1,N\\
		2\csc \xi  & ,\;j=m\\
		-2\cot \xi +\csc \xi  & ,\;j=m-1,m+1\\
		2\tan\left(\xi/2\right) & ,\;\text{other}
	\end{cases},
	\label{analytical solution}
\end{equation}
where $N\geq 5$. From this, one can derive the population $\tilde{P}_{m}$ of the undriven atom, yielding the expression
\begin{equation}
	\dfrac{2\csc^{2} \xi }{1-4\cot \xi \tan\left(\frac{\xi }{2}\right) +3\csc^{2} \xi +( 2N-9)\tan^{2}\left(\frac{\xi}{2}\right)}.
\end{equation}
In the limiting case, we obtain a saturation value $\lim_{\xi \to 0 \;(2\pi)} \tilde{P}_{m} = 2/3$ regardless of $N$, which aligns excellently with the numerical output. Similarly, we have $\lim_{\xi \to 0\;(2\pi)} \tilde{P}_{m\pm 1} = 1/6$. In fact, these results hold true for any $\theta\in(0,\pi)$, revealing the robustness of saturated population against varying incident angles. In other words, the population distribution becomes fully concentrated at the interface atoms, predominantly at the undriven atom as $\xi \to 0$ (or $2\pi$),  leading to pronounced single-site localization. In addition, we provide the interatomic distances, denoted as $\xi_{\max}$, at which $\tilde{P}_{m}$ reaches its maximum for various system sizes $N$. The result can be expressed as
\begin{equation}
	\xi_{\max}=2k\pi +2\tan^{-1}\left(\dfrac{1}{\sqrt{4N-13}}\right),
\end{equation}
where $k\in\mathbb{Z}$. This $\xi_{\max}$ clearly approaches 0 (or $2k\pi$) as $N$ increases, a trend that perfectly matches the observations in Fig.~\ref{fig:figure7}(c).

In summary, as $N$ increases, the maximum achievable population at the undriven atom gradually decreases and converges to the saturation value of $2/3$. Correspondingly, the optimal interatomic distance also approaches $2k\pi$. This phenomenon can be attributed to the competition between the number of atoms and the spin-exchange interactions. As the system size expands, the driven atoms with increasing number carve the population of the undriven atom up and redistributes it into other atomic sites, diminishing the interfaced localization. However, under the influence of strong RDDI (when $\xi \to 0$ or $2\pi$), the population of the undriven atom remains saturated. In contrast, in smaller systems, the population at the interface atoms can slightly exceed the saturation threshold. For example, with $N = 5$, we observe $\max(\tilde{P}_{m})\approx 0.723>2/3$, with corresponding $\tilde{P}_{1, N} \approx 0.092$, and $\tilde{P}_{\text{others}} \approx 0.044$. This observation reflects the subtle competition between system size and RDDI strength within chirally coupled atomic-waveguide systems.

\subsection{Precise Localization Control}
\begin{figure}[htbp]
	\centering
	\includegraphics[width=8.2cm]{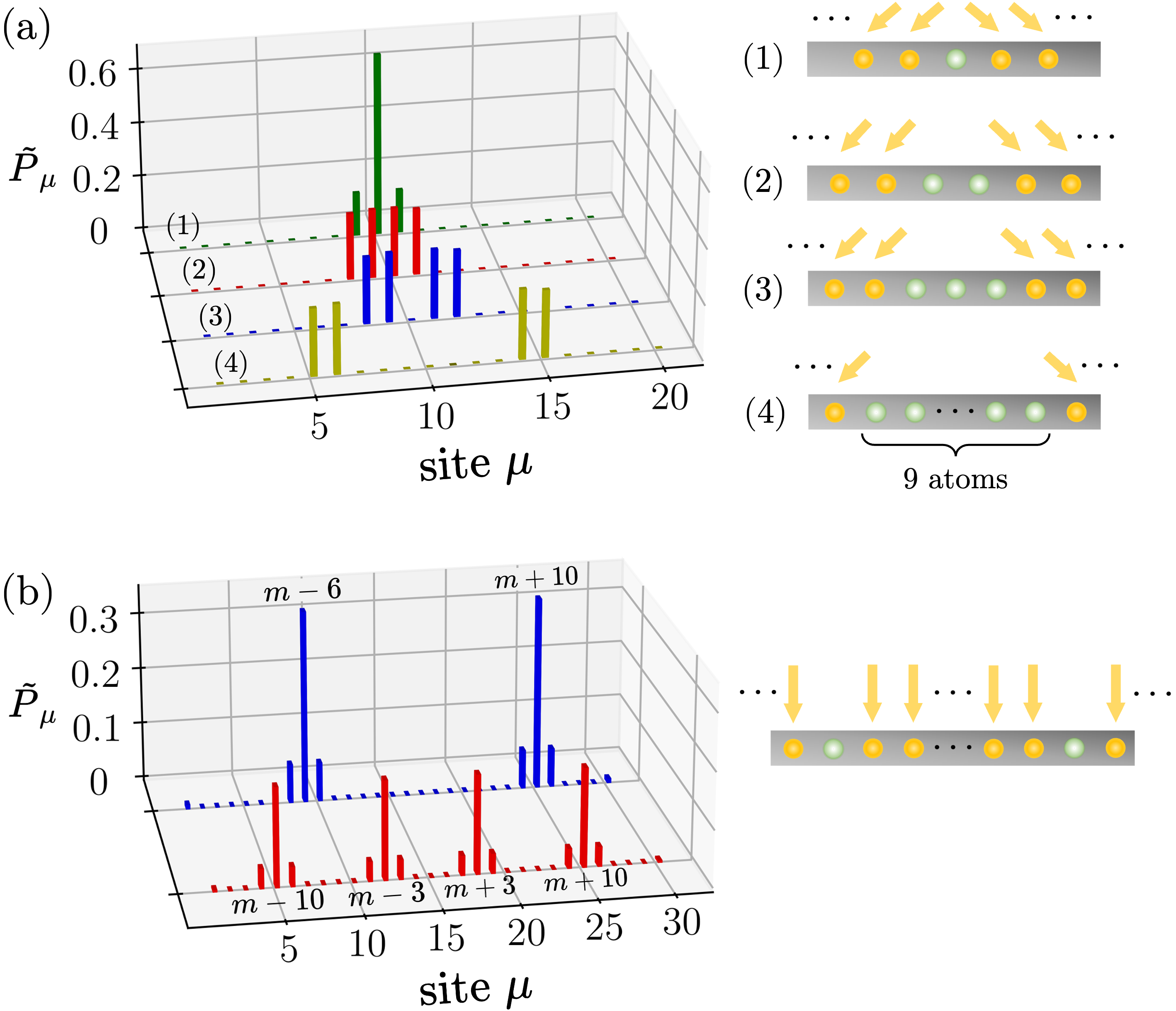}
	\caption{Precise control of localization under the defect-driven scheme. (a) shows the case of $N=20$ atoms under the same parameters in Fig.~\ref{fig:figure7}(b), where removing the driving field from (1) a single atom, (2) two atoms, (3) three atoms, and (4) nine atoms inside (with non-driven atoms shown in green) results in accumulated populations at the two atoms at the ends of the non-driven array and their neighboring sites. (b) demonstrates a method of removing the driving field from several non-adjacent atoms. For $N=30$ atoms, two-site localization depicted in the blue bars is achieved by removing the driving from $(m-6)$-th and $(m+10)$-th atoms, while the red bars illustrate four-site localization resulting from excluding the driving from four non-adjacent atoms ($m-10$, $m-3$, $m+3$, and $m+10$), with $m=15$.}
	\label{fig:figure8}
\end{figure}

By selectively excluding the excitation of a single atom, we induce a robust single-site localization, with the flexibility to reposition this localized excitation by designating different undriven atoms, which offers a framework for precise localization control. This leads to a natural question: what if multiple atoms are left undriven under the same angular configuration? The results, depicted in Fig.~\ref{fig:figure8}, explore this scenario. Specifically, in Fig.~\ref{fig:figure8}(a), we analyze a one-dimensional array of 20 atoms, driven under the same parameters in Fig.~\ref{fig:figure7}(b). Intriguingly, when the excitation field is removed from two adjacent atoms (the 10th and 11th atoms in this instance), the population distributes evenly across these two undriven atoms and their neighboring driven atoms, as illustrated by the red bars in Fig.~\ref{fig:figure8}(a). This configuration effectively truncates population, isolating it to these four atoms. Upon extending the removal to three or even four neighboring atoms, we observe a progressive separation of the distribution, with localization occurring solely at the endpoints of the undriven cluster and their adjacent driven atoms, establishing a controllable approach to excitation localization. Finally, in Fig.~\ref{fig:figure8}(b), we demonstrate an alternative approach for structuring localization distributions. By removing the excitation from non-adjacent atoms in a one-dimensional array, the precise multiple-site localization can be achieved, potentially offering a practical route toward quantum memory implementation.

\section{Conclusion}
\label{sec:conclusion}
In conclusion, a one-dimensional open system of atomic arrays coupled to a waveguide under weak driving forms a distinctive driven-dissipative quantum system capable of exhibiting rich non-equilibrium dynamics and steady-state phases. This arises from the intricate interplay among the competition between long-range dipole-dipole interactions and coupling directionality. By adjusting the incident angles of laser fields, traveling phase differences are introduced, effectively modulating the system's behaviors. We divided the homogeneous array into two subchains, each driven asymmetrically at different angles. This approach, combined with tailored Rabi frequency phase settings, produced strongly localized population distributions. Notably, under the reciprocal coupling regime, adjusting the system size $N$ and atomic separation $\xi$ to satisfy $N\xi \to 2\pi$ enables strongly localized two-site interfaced atomic states with maximal observation probability across various angle combinations. Additionally, precise control over the spatial positioning of this two-site localization can be realized by only shifting the position of the interface between the two subchains, which creates a mechanism to excite population to specific atomic sites. We further redesigned this scheme by removing the laser field on one interfaced atom, designating it as the boundary between the two subchains as a defect-driving scheme. By adjusting the phase of the Rabi frequency, we achieved a more symmetric effective traveling phase. Under these conditions and at a interpartical spacing close to a period of $2\pi$, the saturated population can be observed at the undriven atom, yielding the strongest single-site localization in this study with notable robustness against non-guided decay. This outcome is well-supported by analytical solutions under reciprocal coupling conditions. Our results sheds light on driven-dissipative quantum systems with nonreciprocal coupling, paving the way for quantum simulations of exotic many-body states—a development with significant implications for quantum information applications.

\begin{acknowledgments}
We acknowledge support from the National Science and Technology Council (NSTC), Taiwan, under the Grant No.~NSTC-112-2119-M-001-007 and Grant No.~NSTC-112-2112-M-001-079-MY3, and from Academia Sinica under Grant AS-CDA-113-M04. We are also grateful for support from TG 1.2 of NCTS at NTU.
\end{acknowledgments}





\bibliography{apssamp}

\end{document}